\documentclass[11pt]{article}

        \usepackage{iftex}
        \ifPDFTeX
          \PackageError{sneaking}{This manuscript requires XeLaTeX or LuaLaTeX}{Compile with xelatex or lualatex.}
        \else
          \usepackage{fontspec}
          \usepackage{unicode-math}
          \defaultfontfeatures{Ligatures=TeX,Scale=MatchLowercase}

          \IfFileExists{LibertinusSerif-Regular.otf}{%
            \setmainfont{LibertinusSerif-Regular.otf}[
              BoldFont=LibertinusSerif-Bold.otf,
              ItalicFont=LibertinusSerif-Italic.otf,
              BoldItalicFont=LibertinusSerif-BoldItalic.otf
            ]
          }{%
            \setmainfont{texgyretermes-regular.otf}[
              BoldFont=texgyretermes-bold.otf,
              ItalicFont=texgyretermes-italic.otf,
              BoldItalicFont=texgyretermes-bolditalic.otf
            ]
          }
          \IfFileExists{LibertinusSans-Regular.otf}{%
            \setsansfont{LibertinusSans-Regular.otf}[
              BoldFont=LibertinusSans-Bold.otf,
              ItalicFont=LibertinusSans-Italic.otf,
              BoldItalicFont=LibertinusSans-Bold.otf,
              BoldItalicFeatures={FakeSlant=0.2}
            ]
          }{%
            \setsansfont{texgyreheros-regular.otf}[
              BoldFont=texgyreheros-bold.otf,
              ItalicFont=texgyreheros-italic.otf,
              BoldItalicFont=texgyreheros-bolditalic.otf
            ]
          }
          \IfFileExists{LibertinusMono-Regular.otf}{%
            \setmonofont{LibertinusMono-Regular.otf}[
              BoldFont=LibertinusMono-Regular.otf,
              ItalicFont=LibertinusMono-Regular.otf,
              BoldItalicFont=LibertinusMono-Regular.otf
            ]
          }{%
            \setmonofont{texgyrecursor-regular.otf}[
              BoldFont=texgyrecursor-bold.otf,
              ItalicFont=texgyrecursor-italic.otf,
              BoldItalicFont=texgyrecursor-bolditalic.otf
            ]
          }
          \IfFileExists{LibertinusMath-Regular.otf}{%
            \setmathfont{LibertinusMath-Regular.otf}[BoldFont=LibertinusMath-Regular.otf]
          }{%
            \setmathfont{latinmodern-math.otf}
          }
        \fi

        \usepackage[a4paper,margin=1in]{geometry}
        \usepackage{microtype}
        \usepackage{setspace}
        \usepackage{csquotes}
        \usepackage{booktabs}
        \usepackage{longtable}
        \usepackage{array}
        \usepackage{calc}
        \usepackage{graphicx}
        \usepackage[table]{xcolor}
        \definecolor{UniMelbBlue}{HTML}{033F85}
        \usepackage{caption}
        \newif\ifsneakingrefcallouts
        \sneakingrefcalloutstrue
        \ifdefined\HCode
          \sneakingrefcalloutsfalse
        \fi
        \ifsneakingrefcallouts
          \usepackage{bigfoot}
          \usepackage{etoolbox}
          \makeatletter
          \renewcommand{\footnoterule}{%
            \kern -3pt
            \hrule width \linewidth height 0.4pt
            \kern 4pt
          }
          
          \newlength{\sneakingfnmarkwidth}
          \newlength{\sneakingfngap}
          \newlength{\sneakingfnindent}
          \renewcommand{\@makefntext}[1]{%
            \noindent
            \hangindent=\sneakingfnindent
            \hangafter=1
            \makebox[\sneakingfnmarkwidth][r]{\@textsuperscript{\normalfont\@thefnmark}}%
            \hspace{\sneakingfngap}%
            \emergencystretch=5em\relax
            #1%
          }
          \makeatother
          \DeclareNewFootnote{B}
          \renewcommand{\thefootnote}{\alph{footnote}}
        \fi
        \usepackage{hyperref}
        \hypersetup{
          colorlinks=true,
          linkcolor=UniMelbBlue,
          citecolor=UniMelbBlue,
          urlcolor=UniMelbBlue,
          filecolor=UniMelbBlue,
          pdftitle={From Phreaking to Sneaking: Children's Circumvention of Social Media Age Verification Systems},
          pdfauthor={Bjørn Nansen, Helena Sandberg, Lauren Bliss, Shaanan Cohney}
        }
        \usepackage[user]{zref}
        \usepackage{zref-clever}
        \usepackage[backend=biber,style=authoryear,maxcitenames=2,maxbibnames=99,giveninits=true,uniquename=false,hyperref=auto,backref=true]{biblatex}
        \ifsneakingrefcallouts
        \makeatletter
        \AtBeginDocument{\global\let\sneaking@lastfirstnote\@empty}
        \newcommand{\sneaking@callouttarget}[1]{sneakingcalloutnote:#1}
        \newcommand{\sneaking@calloutanchor}[1]{%
          \Hy@raisedlink{\hyper@anchorstart{\sneaking@callouttarget{#1}}\hyper@anchorend}%
        }
        \newcommand{\sneaking@calloutlink}[1]{%
          \hyperlink{\sneaking@callouttarget{#1}}{#1}%
        }
        \newif\ifsneaking@firstcallout
        \newcommand{\sneaking@calloutseparator}{\textsuperscript{,}}
        \newcommand{\sneaking@emitcallout}[1]{%
          \ifcsdef{sneaking@seen@#1}{%
            \edef\sneaking@thisfirstnote{\csuse{sneaking@firstnote@#1}}%
            \ifx\sneaking@thisfirstnote\sneaking@lastfirstnote
              \footnoteB{\emph{Id.}}%
            \else
              \footnoteB{\citeauthor{#1}, supra note~\sneaking@calloutlink{\sneaking@thisfirstnote}.}%
            \fi
          }{%
            \footnoteB{\sneaking@calloutanchor{\number\value{footnoteB}}\fullcite{#1}}%
            \csxdef{sneaking@seen@#1}{1}%
            \csxdef{sneaking@firstnote@#1}{\number\value{footnoteB}}%
            \edef\sneaking@thisfirstnote{\number\value{footnoteB}}%
          }%
          \xdef\sneaking@lastfirstnote{\sneaking@thisfirstnote}%
        }
        \newcommand{\sneaking@emitcalloutcluster}[1]{%
          \ifsneaking@firstcallout
            \sneaking@firstcalloutfalse
          \else
            \sneaking@calloutseparator
          \fi
          \sneaking@emitcallout{#1}%
        }
        \newcommand{\citecallout}[1]{%
          \begingroup
          \sneaking@firstcallouttrue
          \forcsvlist{\sneaking@emitcalloutcluster}{#1}%
          \endgroup
        }
        \makeatother
        \let\sneakingorigautocite\autocite
        \let\sneakingorigautocites\autocites
        \let\sneakingorigtextcite\textcite
        \RenewDocumentCommand{\autocite}{ O{} O{} m }{\sneakingorigautocite[#1][#2]{#3}\citecallout{#3}}
        \RenewDocumentCommand{\autocites}{ O{} O{} m }{\sneakingorigautocites[#1][#2]{#3}\citecallout{#3}}
        \RenewDocumentCommand{\textcite}{ O{} O{} m }{\sneakingorigtextcite[#1][#2]{#3}\citecallout{#3}}
        \NewDocumentCommand{\fncite}{ O{} O{} m }{\autocite[#1][#2]{#3}}
        \NewDocumentCommand{\fntextcite}{ O{} O{} m }{\textcite[#1][#2]{#3}}
        \else
        \newcommand{\citecallout}[1]{}
        \NewDocumentCommand{\fncite}{ O{} O{} m }{\autocite[#1][#2]{#3}}
        \NewDocumentCommand{\fntextcite}{ O{} O{} m }{\textcite[#1][#2]{#3}}
        \fi
        \newenvironment{interviewquote}{%
          \begin{quote}
          \small
          \setlength{\parindent}{0pt}
          \setlength{\parskip}{0.45\baselineskip}
        }{%
          \end{quote}
        }

        \definecolor{TitleLinkText}{HTML}{111111}
        \definecolor{TitleEmailText}{HTML}{333333}
        \definecolor{TitleLinkCue}{HTML}{666666}
        \IfFileExists{DejaVuSans.ttf}{%
          \newfontfamily\titleiconfont{DejaVuSans.ttf}[Scale=MatchLowercase]
          \newcommand{\titlelinkcue}{\,\textcolor{TitleLinkCue}{\raisebox{0.25ex}{\scalebox{0.62}{{\titleiconfont\symbol{"2197}}}}}}%
        }{%
          \newcommand{\titlelinkcue}{\,\textcolor{TitleLinkCue}{\raisebox{0.25ex}{\scalebox{0.62}{$\nearrow$}}}}%
        }
        \newcommand{\titlelink}[2]{\href{#1}{\textcolor{TitleLinkText}{#2}\titlelinkcue}}
        \newcommand{\titleemail}[1]{\href{mailto:#1}{\textcolor{TitleEmailText}{#1}}}
        \makeatletter
        \newcommand{\titlepagenote}[1]{%
          \begingroup
          \renewcommand{\thefootnote}{}%
          \renewcommand{\@makefntext}[1]{\noindent\footnotesize ##1}%
          \footnotetext{#1}%
          \addtocounter{footnote}{-1}%
          \endgroup
        }
        \makeatother

        \title{From Phreaking to Sneaking: Children's Circumvention of Social Media Age Verification Systems}
        \author{\titlelink{https://findanexpert.unimelb.edu.au/profile/133029-bjorn-nansen}{Bjørn Nansen}\\{\small School of Computing and Information Systems, The University of Melbourne}\\{\small \titleemail{nansenb@unimelb.edu.au}}
\and
\titlelink{https://portal.research.lu.se/en/persons/helena-sandberg/}{Helena Sandberg}\\{\small Department of Communication, Lund University, Sweden}\\{\small \titleemail{Helena.Sandberg@iko.lu.se}}
\and
\titlelink{https://scholar.google.com/citations?user=F094QDAAAAAJ\&hl=en}{Lauren Bliss}\\{\small School of Computing and Information Systems, The University of Melbourne}\\{\small \titleemail{Lauren.bliss@unimelb.edu.au}}
\and
\titlelink{https://cohney.info/}{Shaanan Cohney}\\{\small School of Computing and Information Systems, The University of Melbourne}\\{\small \titleemail{shaanan@cohney.info}}}
        \date{}

\begin{document}
        \maketitle
        \titlepagenote{All authors were equally and collaboratively involved in all aspects of the research paper, including conception and design, data collection and analysis, drafting and revision, critical input, and approval of the final submitted version; all authors agree to be accountable for all aspects of the work.}
        \begin{center}
          \vspace{-0.75\baselineskip}
          {\color{UniMelbBlue}\rule{0.74\linewidth}{0.6pt}}
        \end{center}
        \vspace{0.75\baselineskip}

        \begin{abstract}
        \noindent Australia's social media ban is now in force. It requires platforms to take reasonable steps to stop users under 16 from holding accounts. Drawing on five focus groups with fifteen young people aged 12--16, this paper examines how children understood the ban's effectiveness, impact, and legitimacy as they encountered the platforms charged with enforcing it. Participants widely saw the ban as unfair and ineffective. Through platform access controls, they learned how the ban worked, where it failed, and how they and their peers could evade it. We also asked participants to imagine better approaches to age verification and youth digital governance.

        This paper develops \emph{sneaking} as a theoretical lens for these practices. The concept names more than evasion: it captures the social encounter between children, platforms, techno-regulation, and the access controls that mediate digital participation. Our findings show that children are not passive subjects of platform regulation. They interpret, test, and negotiate digital infrastructure. They also expose a central weakness in age-based platform regulation: technological controls struggle to solve the social and governance problems they are asked to contain.
        \end{abstract}

        \clearpage
        \section{Introduction}\label{introduction}

In late 2025, Australia became the first country to ban social media for users under 16. This paper examines how children directly affected by the ban responded. Drawing on focus groups with children aged 12--16, we ask how they viewed, navigated and, in many cases, circumvented, the technologies implementing Australia's social media restrictions.~Our participants broadly disagreed with the ban; and frequently bypassed it. The participants described both uneven platform enforcement and their tactics and rationales for keeping access to social media. To theorise these practices, we develop the concept of social network \emph{sneaking --} a concept which draws on the technical and cultural legacy of phreaking. We use this framing to analyse children's development of a culture of bypassing age verification systems, while also suggesting its broader relevance for understanding how users work around digital restrictions generally. This lens also emphasises children's rights in digital environments \autocite{livingstoneOneill2014,unCommittee2021}, especially their rights to information provision, participation, and protection.

We identify four themes that capture our participants' approach to sneaking: 1) young users learn through experimentation how age verification systems work; 2) they manipulate systems to bypass restrictions; 3) they share vernacular knowledge about vulnerabilities; 4) and they imagine and attempt to contribute to alternative futures. Across these themes, children's circumvention appears less as an overt challenge to authority than as a pragmatic negotiation with digital infrastructure -- much like the early culture of ``hackers''. This connects sneaking to concerns in studies of infrastructure politics about platform governance and the limits of technical enforcement, which we return to in the discussion. Young people's response to existing systems is more than passive. Children's everyday workarounds represent an emerging engagement with the politics of digital infrastructure, which we argue that this demonstrates their capacity to imagine fairer forms of digital governance. In our framework, young users are not merely subjects of control but rights-bearing agents who interpret, navigate, and work around systems they perceive as unjust or misaligned with their lives.

\section{Background and Context}\label{background-and-context}

\subsection{The Australian Age Restrictions and Age Verification Technologies}\label{the-australian-age-restrictions-and-age-verification-technologies}

Australia's recent restrictions on children's access to social media platforms are a world-first: no prior law had imposed on platforms a \emph{positive obligation to take reasonable steps}, typically through age assurance measures, to prevent under-16s from holding accounts \autocite{champion2025,esafety2026a,onlineSafety2024}. The restrictions took effect on December 10, 2025 \autocite{departmentInfrastructure2025a}.

Children's own experiences of the ban are an especially important and thus far understudied phenomenon. Although extensive research examines children's online safety and media use \autocites[e.g.,][]{gathSwit2024,smahel2020,stoilova2021}, much less empirical work has examined how young people encounter age verification technologies or how abrupt policy shifts reshape their access to, and practices around, social media. This matters beyond Australia because the ban is already being explicitly cited in active policy and legislative processes elsewhere, including a UK government consultation on minimum age for social media access, a French parliamentary bill to ban social media for under-15s, and German proposals for age-based social media curbs \autocite{departmentScience2026,reuters2026a,reuters2026b}. Through youth-centred qualitative methods, this paper shows how regulation, platform infrastructure, and children's agency interact, and contributes to debates about the ethical design of verification systems and youth protection policy in digital environments.~

In practice, compliance is relying on a range of age assurance mechanisms, including document-based checks (credit cards/IDs), machine-learning systems that estimate age from a photo or video of a user's face, and systems that infer a users' age based their platform use. Among these, facial age estimation systems have attracted particular public attention. Unlike document-based checks, these systems are designed to estimate whether a user falls within a given age range rather than verify their identity. This reduces reliance on government-issued identification, but introduces a different set of technical and normative concerns.~State-of-the-art facial age-estimation systems still produce non-trivial false-positive and false-negative errors, especially in age bands close to relevant policy thresholds. Further NIST evaluations of face-analysis systems have also documented demographic performance differentials, including skin-tone-related effects in some error patterns \autocite{hanaoka2024,grother2022}.~The models also raise distinct privacy concerns \autocite{departmentInfrastructure2025b,hfAn2025,kohler2025,lueks2026}.

The policy is controversial not only because of the technologies on which it may depend, but also because of Australia's policy design and implementation. Critics point to uneven platform coverage, the chosen age threshold, rushed passage of the law, gaps in the robustness of implementations, and unintended consequences for both child and adult users \autocite{esafety2026a,taylor2026,rubinstein2024smma}. These implementation concerns also expose a deeper policy tension: how to protect children while respecting their developing autonomy and right to inclusion in digital life \autocite{stalfordLundy2025}.~Children's involvement in decisions about digital policies matter because their everyday social, educational, and civic practices are now deeply embedded in digital environments \autocite{kohler2025,livingstoneThird2017}.

Whether these measures are justified depends in part on whether they can deliver their stated benefits in practice. Before implementation, public reporting had already identified VPNs, AI-generated images, and disguise-based attempts as likely bypass vectors \autocite{lavoipierre2025}.~Yet eSafety's March 2026 compliance update and subsequent reporting \autocite{butler2026} indicated that many children retained or regained accounts. Similarly, eSafety's own survey of parent's suggested that around seven in ten children with accounts on several major platforms still had access \autocite{esafety2026b,taylor2026}. This is consistent with our findings. Our young participants described age verification as easy to circumvent. These accounts sit uneasily alongside a government-commissioned report, prepared by an industry partner from the age assurance industry, which claimed that age assurance could be implemented privately, efficiently, and effectively \autocite{departmentInfrastructure2025b}.

Drawing on youth-centred qualitative methods this paper offers an original account of the interplay between regulation, platform infrastructure, and children's agency. This contributes directly to the ethical design and implementation of emerging verification systems and the refinement of youth digital protection policies. The discussion suggests that children may not simply encounter the ban as a rule to obey or resist. Rather, the interaction is an encounter with a technical system to be interpreted, tested, and worked around. In the following subsection, we develop a lens to enrich our discussion of the interaction's collective, knowledge-sharing, and infrastructural dimensions.

\subsection{Sneaking: A New Theoretical Framework}\label{sneaking-a-new-theoretical-framework}

To make sense of children's patterned practices of interpreting, testing, and bypassing age verification systems, we develop \emph{sneaking} as a conceptual lens. Sneaking adapts the technical and cultural lineage of \emph{phreaking} to the context of children's interactions with social media age verification \autocite{lakoffJohnson1980,ludlow2022,warnick2004}. Our construction of sneaking helps us analyse how children learn the operational logics of access systems, identify weak points, and circulate practical knowledge about how those systems can be worked. To clarify what this borrowing captures, it is useful to begin with phreaking itself.

Phreaking, a portmanteau of ``phone'' and ``freak'' (1960s slang for unorthodox or free), emerged in the 1960s to describe the act, or art, of exploring, manipulating, and hacking telecommunications systems to make free phone calls or access restricted services \autocite{levy1984}. Phreaks identified audio vulnerabilities in public telecommunications networks and developed techniques for using analogue signal tones to mimic operator controls. By emitting a 2,600 Hz tone into a phone, they could make the system read the line as open and route calls for free. A youth subculture grew around this practice. Young people shared tips such as using the toy whistle from Cap'n Crunch cereal boxes, which produced the right frequency; published technical information in newsletters such as YIPL/TAP (Youth International Party Line / Technological Assistance Program); and made devices such as the ``blue box,'' later associated with Steve Wozniak and Steve Jobs \autocite{draperFraser2018}.

Phreaking declined in the 1980s after digital switching, including Common Channel Interoffice Signalling (CCIS), separated signalling from voice lines and made the classic tones obsolete \autocite{gold2011}. Even so, phreaking is widely understood to have helped lay the technical and cultural foundations for computer hacking after personal computers and microcomputers developed in the 1970s and 1980s \autocite{levy1984}. Phreaking shaped technical curiosity about opening and jailbreaking hardware and software systems. It also informed open-source ideologies and anti-corporate or anti-government critiques of computing and internet infrastructure, grounded in values of experimentation, discovery, tinkering, community, and collective knowledge sharing \autocite{coleman2012,levy1984}. The practice was fundamentally playful. Although phreakers circumvented systems for practical and ideological reasons, they often did so pragmatically, seeking personal ends and shared cultural practice rather than explicit collective protest.

Building on this legacy, we coin the concept of sneaking, derived from the terms ``social network freaking.'' We extend the historic subcultural practice of phreaking as a conceptual lens to understand how young users perceive, interpret, and act within digital systems. Moreover, the ordinary verb to sneak names stealthy, covert, or secretive actions that avoid detection, including actions used to bypass age verification. And so, as we discuss below, the concept of sneaking carries implicit meaning and signifying capacity for thinking through emergent forms of user navigation of digital infrastructures and their governance. Like phreaking, sneaking points to personal and cultural practices of navigating and bypassing communication or security systems, here social media age verification. At a technical level, age verification systems govern access by requiring users to prove their age before entering. Much like older telephone systems, these systems have weak points, inconsistencies, and informal bypasses. Young people learn those bypasses through everyday use and develop informal expertise about system signals and operational logic. Sneaking therefore foregrounds the social tactics of discovering, developing, and distributing infrastructural knowledge.

In the context of age verification, sneaking is not a direct challenge to authority. It is a set of practices and techniques for finding vulnerabilities, evading or exploiting rules, and achieving a desired personal outcome: continued access to social media accounts. Conceptually, this aligns with materialist media theory, which treats hacking and vulnerability exploitation as foundational to contemporary computational, communication, and security apparatuses \autocite{chun2011,gillespie2018,kittler1999,pasquale2015}. \textcite{galloway2004} and \textcite{gallowayThacker2007} argue that digital networks replace older hierarchical or disciplinary forms of power based on institutional authority \autocite{foucault2018} with control \autocite{deleuze2017}: power operates through coded systems and technical protocols that govern communication. In this framework, sneaking names technical and cultural practices that enable continued digital participation by navigating access protocols such as age verification. These practices reflect children's efforts to realise their rights and desire to participate by identifying and negotiating limits embedded in platform infrastructures and governance.

\section{Research Methods}\label{research-methods}

This study examined how children understood and responded to the restrictions in their immediate wake. Because the ban directly affected children, their accounts are central to understanding its impact. Focus groups capture contextualised and dialogic perspectives on technology use and experience \autocite{morgan1997,kitzinger1994}. They also show how children negotiate shared understandings, co-construct narratives, and challenge each other's assumptions \autocite{hennessyHeary2005}. Peer-to-peer discussion is especially valuable for studying relational technologies such as social media, which operate through social expectations and group dynamics. Focus groups can also elicit examples that participants may not state publicly, including unspoken rules about what is acceptable.

We conducted five in-person focus groups with children aged between 12-16 to capture their expertise and contextualised, dialogic perspectives \autocite{hennessyHeary2005}. The focus groups took place between December 2025 and February 2026, immediately after Australia's social media age restrictions came into effect on December 10, 2025 \autocite{departmentInfrastructure2025a}. Each group lasted up to one hour and included two to four participants, with a total of 15 participants (see \zcref[cap]{tab:participant-demographics}). We stratified groups by age (e.g., 12-14, 14-16) to support developmental appropriateness. Children were recruited through posts in school newsletters, with focus groups held at participant-chosen location. Parents or guardians provided consent, and children provided assent. The study was approved by out University ethics committee (removed for review).

\begin{center}
\begin{minipage}{\linewidth}
\captionsetup{type=table,hypcap=false}
\centering
\caption{Participant Demographics}
\zlabel{tab:participant-demographics}
\footnotesize
\setlength{\tabcolsep}{0.55em}
\renewcommand{\arraystretch}{1.05}
\begin{tabular}{@{}lcl@{\hspace{1.8em}}lcl@{\hspace{1.8em}}lcl@{}}
\toprule
ID & Age & Gender & ID & Age & Gender & ID & Age & Gender \\
\midrule
P1 & 13 & Female & P6 & 12 & Female & P11 & 13 & Female \\
P2 & 14 & Female & P7 & 13 & Female & P12 & 16 & Male \\
P3 & 15 & Male   & P8 & 16 & Female & P13 & 15 & Male \\
P4 & 14 & Male   & P9 & 14 & Male   & P14 & 15 & Male \\
P5 & 12 & Female & P10 & 15 & Male  & P15 & 15 & Male \\
\bottomrule
\end{tabular}
\end{minipage}
\end{center}

We organised the focus group discussions around four topics: (1) how children understood and perceived the social media restrictions and age verification technologies; (2) how they negotiated banned and approved social media platforms; (3) how they used digital platforms after the restrictions; and (4) how the ban affected their social lives, friendships, and wellbeing. The focus groups also included a participatory co-design activity in which children creatively and critically reimagined alternatives to the current social media age restrictions \autocite{hourcade2023,valguarneraLandoni2023}. The activity moved beyond eliciting opinions by involving young participants in shaping ideas for real-world policy implementation. This approach positions children as active contributors and knowledge producers rather than passive subjects and aligns with ethical, rights-based research principles \autocite{howard2025,livingstoneThird2017,sandbergGille2021}.

We analysed the focus group transcripts using thematic analysis \autocite{braunClarke2006}. We imported transcripts into Delve, a qualitative data analysis platform, to organise, code, and retrieve textual data and to develop analytic categories. A grounded theory-oriented coding strategy informed the analysis \autocite{corbinStrauss1990}. First, we used open coding to identify and label salient concepts and patterns in the data. We then used axial coding to compare, organise, and synthesise codes into higher-order categories, with attention to relationships between themes. To support consistency in interpretation and reflexive engagement with the data, coding decisions and the development of thematic categories were reviewed and discussed by the authors.

\section{Findings}\label{findings}

Our key findings capture four themes. First, young users explored and learned about age verification systems and vulnerabilities, treating checks and bans as navigable technical rules rather than absolute barriers. Second, they manipulated system requirements to bypass restrictions. Third, they developed shared vernacular knowledge about vulnerabilities in digital networks and verification systems. Fourth, they used design imaginaries to co-design alternative futures for more effective and equitable age verification. Each theme is presented below and supported by examples from our data.

\subsection{Age Verification as Platform Access Control Protocols}\label{age-verification-as-platform-access-control-protocols}

This theme describes how children anticipated, experienced, explored, and understood age verification mechanisms. The analysis shows that children saw platforms as the main gatekeepers for age verification and enforcement. They understood the policy through specific technical protocols, especially date-of-birth confirmation and selfie-based facial age-estimation checks. We organise this theme around two categories: children's views and experiences of platform enforcement, and their familiarity with security vulnerabilities based on prior experiences of poorly implemented social media rules.

\subsubsection{Platform Enforcement}\label{platform-enforcement}

The social media restrictions were widely publicised, and children were acutely aware of their impending implementation. Closer to the implementation date, children received notifications to verify their age or risk losing accounts on several platforms, including the two dominant platforms discussed in the research: Snapchat and TikTok:

\begin{interviewquote}
\emph{``Yeah, on Snapchat, it like gave you a notification. Like, every few days it would come back,} \emph{and it was like, verify your age now, or your thing will get banned'' (P5, female 12)}

\emph{``I had a little notification, like, in three days your account could be getting banned from the} \emph{age thing. And then three days passed, I still have my account'' (P3, male 15)}
\end{interviewquote}

Despite these warnings, children expressed little concern about the impact. They thought the ban was ineffective based on their initial experiences and its uneven enforcement. Many participants described not being verified at all. Others described warnings that enforcement was imminent, or cases where access was later restored:

\begin{interviewquote}
\emph{``Every single person I know that has gotten banned on Snapchat, they got their account back,} \emph{like, after like three days'' (P3, male 15)}

\emph{``Nothing happened to me at all. \ldots{} I think for TikTok, I had my real age, but I {[}still{]} wasn't} \emph{affected at all. I'm not really sure how {[}but{]} it is like a world's first, so you kind of, I wouldn\textquotesingle t} \emph{expect it to be perfect'' (P13, male 15)}

\emph{``I got banned on my Snapchat account. But then I tried to log back in, like, a week later and it} \emph{just let me in'' (P7, female 13)}

\emph{``Oh, so you were banned on the actual day?'' (Researcher)}

\emph{``Yeah, like for a couple of days, but I just logged in and was let me back in'' (P7, female 13)}
\end{interviewquote}

Many children also felt that social media companies were only superficially enforcing the ban and viewed corporate compliance as minimal, technical, profit-driven, and open to circumvention. They also recognised that platforms could still encourage under-age use by enabling access without an account on platforms like TikTok and YouTube:

\begin{interviewquote}
\emph{``They just do the bare minimum to make it sound like they\textquotesingle re trying'' (P5, female 12)}

\emph{``It\textquotesingle s because they don\textquotesingle t want to lose their consumers'' (P10, male 15)}

\emph{``You can still watch social media without having an account. You don\textquotesingle t need an account to use} \emph{YouTube. I've got TikTok without an account. It just means you can\textquotesingle t, like, comment or} \emph{anything,} \emph{or like anything or post anything'' (P2, female 14)}
\end{interviewquote}

\subsubsection{Security Vulnerabilities}\label{security-vulnerabilities}

Children's knowledge that age checks were based on self-reported birth dates or selfie-based facial age estimation, rather than physical IDs or human checking, helped them see the system as bypassable. They also recognised an asymmetry of accountability in enforcement: platforms were responsible for taking reasonable steps, while children experienced the checks as weak or hollow. Age verification was therefore understood as a technical process open to circumvention and manipulation:

\begin{interviewquote}
\emph{``But I think it\textquotesingle s like a software thing that failed. Like the way that they did it. It\textquotesingle s like matching} \emph{just what someone\textquotesingle s written, with no proof, no evidence that it\textquotesingle s actually their birthday'' (P8,} \emph{female 16)}
\end{interviewquote}

Children's experiences and perceptions of past policies are also crucial to understanding the effectiveness of current age verification systems. Alongside their awareness of technical flaws, children described a gap between age-based platform rules and how those rules worked in practice. Some said they had already lied about age because earlier platform rules restricted children under 13 without parental permission. These prior experiences made lying about age a familiar workaround:

\begin{interviewquote}
\emph{``It's not gonna work, because no one puts down their real date of birth'' (P8, female 16)}

\emph{``They literally just asked me what my age was, and then there was, like, are you sure? I just} \emph{clicked `yes I\textquotesingle m sure' {[}laughs{]}'' (P4, male 14)}
\end{interviewquote}

Restrictions may have failed or been circumvented technically, but children also described cases where verification worked through social norms. Access rules appeared more effective when reinforced by parental authority, perceived legitimacy, or the internalisation of norms:

\begin{interviewquote}
\emph{``I didn\textquotesingle t do it {[}face verification{]} because I didn't know if mum and dad would let me ...'' (P6,} \emph{female 13)}
\end{interviewquote}

\begin{interviewquote}
\emph{``I didn't do it {[}challenge age verification{]} because I knew that I wouldn't `look the part' and that I probably shouldn't push it since obviously the um rules have been put up, like, for a reason'' (P11, female 13)}
\end{interviewquote}

Notably, no participant reported migrating to alternative platforms such as Lemon8, despite media noting this would happen, suggesting that platform value is not easily substitutable. As one participant described, ``there\textquotesingle s no kind of alternate version of TikTok'' (P5, female 12).

\subsection{Young Users Bypassing Platform Access Controls}\label{young-users-bypassing-platform-access-controls}

This theme describes how young users treated age verification systems as infrastructure to be probed, tested, and worked around. In contrast to cases where no verification occurred or checks were not enforced, these accounts show children actively detecting weaknesses and developing bypass tactics. We coded these tactics into three modes: data falsification, infrastructural circumvention, and visual manipulation.

\subsubsection{Data Falsification}\label{data-falsification}

Data falsification involved children deliberately utilising fake personal information to pass age gates. These forms of falsification included established exploits like entering fake birthdates, as noted above, or creating fake accounts. It also involved more responsive data and identity fabrications in response to the age ban, such as borrowing friends' accounts, or using siblings' or parents' details:

\begin{interviewquote}
\emph{``With the verification you can just use like a fake school ID, or you can use like your parents} \emph{ID''} \emph{(P2, female 14)}

\emph{``I know some people use like, people that are older, like their cousins, and it just works'' (P6,} \emph{female 13)}
\end{interviewquote}

The use of false data can be viewed as the simplest and lowest-friction workaround, and something we have noted is already normalised in children's routine digital practices.

\subsubsection{Infrastructural Circumvention}\label{infrastructural-circumvention}

In contrast to falsifying data, children also described tools and tactics for bypassing enforcement systems based on their awareness of platform protocols. These infrastructural techniques included the potential use of VPNs or non-banned social media platforms, although this did not eventuate because children found easier workarounds. Other approaches included account or device switching, resetting accounts after bans, using multiple accounts, and creating new accounts on the same device:

\begin{interviewquote}
\emph{``People could go on their iPad and make a whole new account'' (P2, female 14)}

\emph{``My TikTok account got banned, but I just made a new account, and it works ... I just said I was} \emph{16'' (P7, female 13)}
\end{interviewquote}

Children shared experiences, discoveries, and ideas about creating multiple accounts and using different devices. These included creating a new account with a different birthday, using another household device to make a new account, or drawing on earlier experience with fake accounts. Multiple accounts helped participants reduce the risk of losing access:

\begin{interviewquote}
\emph{``I got banned. But I already had another account, that I'd made in case, and I just couldn't be} \emph{bothered trying to fix {[}the old one{]}'' (P7, female 13)}

\emph{``But everyone, like, just made other accounts with a new fake age'' (P7, female 13)}
\end{interviewquote}

This mode required greater degrees of digital literacy than data falsification, though it still emerged as common amongst our participants.

\subsubsection{Visual Manipulation}\label{visual-manipulation}

A particularly salient and creative mode of bypassing age verification involved manipulating images used by facial age-estimation systems. Participants described developing and applying tactics through trial and testing, rumour, peer knowledge sharing, and online searching. These tactics included asking an older relative or friend to complete age checks, holding an image of an older face to the camera, and having users of different genders and ages complete facial scans for each other:

\begin{interviewquote}
\emph{``Like, my little sister, her Snapchat, like, asked her to verify her age. To face scan. And I just did} \emph{my face, and it was fine. I just put my face in'' (P12, male 16)}

\emph{``I know some people use like, people that are older, like their cousins, and it just works'' (P6,} \emph{female 13)}

\emph{``And you can also do this, like face scans on YouTube, where it's literally a video of, like} \emph{someone doing a face scan, and you just put your phone in front of that'' (P14, male 15)}
\end{interviewquote}

These image manipulation tactics also involved experimenting with makeup, costumes, and lighting. Children realised that lifelike verisimilitude was not necessarily important to AI image reading, so exaggerated or humorous disguises, including fake beards, could be used to appear older even if they would look implausible to a human. Participants described how they and their friend's used creative tactics as makeup, lighting, and costuming to shape how automated systems interpreted their age:

\begin{interviewquote}
\emph{``She did red lipstick'' (P7, female 13)}

\emph{``Someone wore a Santa beard. And that worked'' (P5, female 12)}

\emph{``I did a face scan for Snapchat. I literally just did it in the dark, so it worked'' (P4, male 14)}

\emph{``Is that what your trick was to try to, like, dim the lighting?'' (Researcher)}

\emph{``Yeah. And then I, like, tried to make myself look as old as possible'' (P4, male 14)}

\emph{``By doing what?'' (Researcher)}

\emph{``I don\textquotesingle t even know, looking serious. And I like, sucked in my chubby cheeks'' (P4, male 14)}
\end{interviewquote}

These practices were often recounted with laughter, showing how humour helped render the system legible. Through such exchanges, children translated opaque computational processes into embodied, manipulable features of lighting, texture, and facial cues. This mode was a particularly advanced and creative response as platforms adopted stricter age assurance technologies.

\subsection{Circumvention as Vernacular Knowledge Sharing}\label{circumvention-as-vernacular-knowledge-sharing}

The third theme captures how circumvention operated as a form of digital literacy within children's networks. The practices described above show that sneaking was not only individual bypassing. It was also a collective process of decoding how systems worked through shared cultural knowledge, interpretive tactics, and peer learning. Rather than simply violating rules, children developed a vernacular digital security literacy. They experienced the ban as a cohort who had previously had access, and many responded by refusing, circumventing, and sharing knowledge in order to maintain established forms of participation. We identify two categories within this theme: knowledge sharing and the ideologies underpinning the restrictions.

\subsubsection{Knowledge Discovery and Sharing}\label{knowledge-discovery-and-sharing}

In terms of information discovery and knowledge sharing around circumvention tactics they had found or heard about, children listed a range of sources, including online searching, social media feeds, school and peer-based conversations, and general rumours circulating about effective tactics:

\begin{interviewquote}
\emph{``Um, one of our friends did. And I\textquotesingle ve also, like, heard people online with like, the verification they can use, like, mascara, and they put like, um, like, to make it look, like, they have facial hair. And apparently, if they have any facial hair, it automatically thinks they\textquotesingle re an adult, so I\textquotesingle ve seen people do that'' (P13, male 15)}

\emph{``We definitely get a lot of our information sources from social media'' (P14, male 15)}
\end{interviewquote}

Children also expressed an understanding of the limitations and locus of power in enforcing the age ban, given the sense that it was ineffective. They noted the limits on platforms' ability to adhere to government policy. In contrast, children understood the home as a key site of rule enforcement and often saw parental oversight as more effective than platform-based controls:

\begin{interviewquote}
\emph{``It will be interesting to see if any of them actually get fined, because if they get found that they didn\textquotesingle t take reasonable steps they can get fined \$50 million'' (P11, female 13)}
\end{interviewquote}

\begin{interviewquote}
\emph{``Yeah but what is considered `reasonable steps'? There\textquotesingle s no, like simple answer to what is a reasonable step. Like they went to reasonable steps of putting in new AI to verify the age. They verified people\textquotesingle s ages; if they got around that, that\textquotesingle s on them. It\textquotesingle s not on the social media'' (P10, male 15)}

\emph{``I think parents have more power than, like, the actual government. Like, realistically, the government could ban you from the apps. But say your mum didn\textquotesingle t agree with it, she could literally take your iPad and just delete the app. The government can\textquotesingle t physically, like, get you fully rid of, like, all the apps they want to get rid of, but your parents can'' (P3, male 15)}
\end{interviewquote}

\subsubsection{Restriction Ideologies}\label{restriction-ideologies}

Children recognised the limits of enforcing the restrictions. They recognised that stricter enforcement, such as ID checks or government databases, raises privacy and practicality concerns. Some also interpreted the ban as political or ideological, seeing it as a performative response rather than a fully effective protection measure. This was further expressed in noting the limited number of platforms included in the ban:

\begin{interviewquote}
\emph{``So they're {[}the government{]} just banning Snapchat, and all they're doing is they\textquotesingle re like priding themselves as the first country to do it. It\textquotesingle s so much bullshit. There\textquotesingle s so much politics'' (P9, male 14)}
\end{interviewquote}

\begin{interviewquote}
\emph{``I think it\textquotesingle s kind of weird. Like, they try to ban it, but they don\textquotesingle t ban half the things'' (P7, female 13)}
\end{interviewquote}

\begin{interviewquote}
\emph{``Like, they don\textquotesingle t ban Discord'' (P5, female 12)}
\end{interviewquote}

\begin{interviewquote}
\emph{``Yeah, like they banned only a few of the biggest ones. But like, what\textquotesingle s even the point in trying to make a ban then if there's still half of them?'' (P7, female 13)}
\end{interviewquote}

Taken together, these accounts point to the development of shared vernacular literacy in response to an unexpected and seemingly unfair ban. Children demonstrated an emergent understanding of how systems govern access: what data they require, how verification processes operate, where loopholes might exist, and how this knowledge can be shared.

\subsection{Design Imaginaries for Alternative Verification Systems}\label{design-imaginaries-for-alternative-verification-systems}

Children's imaginaries for improving age verification systems moved beyond critique to speculative design. In the participatory co-design element of the focus groups, children articulated how systems might better align with their lived realities, implicitly proposing different configurations of trust, identity, and participation. This orientation is evident across two dimensions of their accounts: alternative verification measures, and a forward-looking belief about the benefits of the ban for future cohorts who have not yet engaged with these platforms.

\subsubsection{Alternative Verification Measures}\label{alternative-verification-measures}

When asked, ``If you could redesign how age rules work online, what would you change?'' participants suggested institution-based access, such as access through schools rather than birthdays. They also suggested platform-level feature restrictions. They valued limiting social media to peer interaction while curbing broader exposure to content, alongside stronger safety mechanisms to address harmful behaviour. Their proposals emphasised more nuanced, graduated forms of access rather than blanket bans, suggesting that limiting specific features would be more effective than removing platforms altogether:

\begin{interviewquote}
\emph{``I think it should be what grade you\textquotesingle re in at school, because they could introduce it through the schools or something. Yeah, maybe then everyone kind of gets it at the same time instead of your birthday'' (P7, female 13)}

\emph{``And maybe not banning Snapchat completely, yeah. I feel like they can just remove features like spotlight'' (P14, male 15)}

\emph{``It shouldn\textquotesingle t even be like a whole separate app. It should just be within the app. If you\textquotesingle re this age, then it just, like, might restrict, like snap maps, or, you know'' (P12, male 16)}
\end{interviewquote}

\subsubsection{Benefits for Future Generations}\label{benefits-for-future-generations}

Participants framed the ban's impact as uneven across ages and cohorts. It was most disruptive for children already using social media, especially where it involved losing established accounts and social connections. By contrast, participants anticipated that future cohorts might experience the restriction as less consequential or even beneficial if access were delayed from the outset. Some suggested that a universal ban could reduce exclusion and reshape social media norms, while others saw delayed access as a way to preserve childhood and slow premature exposure to online environments:

\begin{interviewquote}
\emph{``I reckon if you already had it, it affects you more, but if you didn\textquotesingle t have it, you wouldn\textquotesingle t really care'' (P6, female 13)}

\emph{``I think that it will, like, kind of give them a minute to still be kids, and then they don\textquotesingle t have to, like, rush into anything and, like, grow up so fast'' (P11, female 13)}

\emph{``I feel like it would work because if it banned everyone then you wouldn't be left out, you know what I mean. Then you would be like, it's, oh fine, I don't actually need it. And I think you would kind of change your perspective on social media, that you don't actually need it'' (P5, female 12)}
\end{interviewquote}

Taken together, these design and age proposals reflect a preference for more context-sensitive and socially grounded forms of governance.

\section{Discussion}\label{discussion}

The historical practice of phreaking is a useful starting point because it names a pattern in which security systems become legible through attempts to restrict access. Our findings show a similar pattern in children's responses to age verification: participants encountered verification as a set of practical systems to be tested, identified gaps in platform access controls, and developed and shared techniques for bypassing them. Three linked practices make up this form of sneaking. First, children encountered date-of-birth checks, inconsistent enforcement, and selfie-based age verification age-estimation systems as navigable rather than absolute. Second, they bypassed and tricked systems using false ages, new accounts, older friends or relatives, videos, lighting, makeup, and costumes to pass facial age estimation. Third, sneaking was socially learned through friends, school conversations, social media feeds, online searching, and rumours about effective tricks. This shared vernacular knowledge turned verification from a private obstacle into a collective technical challenge. Because participants could often circumvent protocols, they tended to view the ban as largely ineffective. Their workarounds, however, were not simply non-compliance. They show why age verification needs to be understood sociotechnically, as a form of infrastructural politics \autocite{acland2015}.

Seen this way, sneaking reframes apparent rule-breaking as testing, decoding, and navigating systems of control from within. This aligns with accounts of digital politics as immanent to networked systems, by finding vulnerabilities within access rules themselves, rather than opposing networks from the outside \autocite{deuze2011,gallowayThacker2007}. But sneaking also complicates exploit theory as an overtly political mode of resistance. These children were not necessarily engaged in explicit resistance or collective protest. They were pragmatically navigating systems that had made their existing social media participation uncertain. Sneaking can therefore be understood as infrastructural politics in that children recognised that platform rules, protocols, biometric systems, and organisational decisions were not neutral. They shaped who could access digital spaces and on what terms. This matters because the policy and technical decisions that governed children's access had not meaningfully included them, producing outcomes they experienced as unfair and unequal. Sneaking bridges intentional resistance and ordinary use: children may seek the practical outcome of continued access, but their collective navigation exposes the limits of technical governance.

While we use sneaking as a conceptual lens for understanding how children develop informal expertise, techniques, and communities of knowledge-sharing in the contexts of social media age verification systems, this remains one form of digital sneaking. We suggest that the concept of sneaking has wider application and significance for understanding a range of tactical forms of encounter, navigation, and deviation from the strategies of power defining digital networks \autocite{deCerteau1984}. The operational power maintained by digital platforms necessitates sneaking as a mode of survival, pragmatism, or politics in a range if online contexts. Just as phreaking emerged as a cultural practice in response to the dominance of large and corporate technological infrastructures, the concept of sneaking both parallels the past, but also signals to the emergence of a more diverse range of cultural practices associated with navigating digital platforms that are organised as a tightly stacked complex of technological, economic, and governance infrastructures structuring our daily lives \autocite{plantinPunathambekar2019,srnicek2017}. Sneaking is furtive, it is playful, it is experimental, and it is collective; it produces knowledge, techniques and empowerment that have potential to reorganise or reimagine the conditions of digital life, such as those of children's digital rights. Age verification protocols encode access and regulate participation through platform-controlled gates. Yet, the child-rights perspective within sneaking changes the interpretation of participants' workarounds: their testing and circumvention point to unmet claims for privacy, access, and meaningful involvement in the governance of digital environments \autocite{livingstoneThird2017}. Sneaking therefore centres children as active agents within these infrastructural systems rather than passive objects in need of protection.

Participants' play, testing, and circumvention also produced vernacular knowledge. Their accounts of fake ages, account switching, peer tips, and facial-estimation tricks show that verification was understood as inputs, thresholds, and outputs that could be manipulated. Consistent with scholarship on vernacular creativity \autocite[e.g.][]{gibbs2015}, this literacy was social, iterative, and collectively maintained. Further, consistent with theories of algorithmic imaginaries and folk understandings of technology \autocite{bucher2017,ytreMoe2021}, these perspectives also reveal an awareness of privacy risks associated with data-driven verification, alongside a capacity to reimagine how such systems might be reconfigured in more accountable and effective ways. The design suggestions extend this argument. Participants' frustrations with blanket bans were linked to alternatives such as institution-based access, parental involvement, and platform-level feature restrictions. If protocols organise participation \autocite{galloway2004}, then children's proposals show an emergent capacity to imagine how those rules might be otherwise structured.

\section{Conclusion}\label{conclusion}

This paper has proposed the concept of sneaking, developed from the history of phreaking, to explain how children navigate social media age verification. The findings show children encountering inconsistent enforcement, testing verification thresholds, sharing workarounds, and proposing alternative forms of access governance. Age verification is therefore not only a technical measure. It is a negotiated digital infrastructure through which values of child trust, equality, privacy, and rights are contested. Across the four themes, children appeared as active rights-bearing agents rather than passive subjects of platform control. They learned how age gates worked, manipulated technical signals to maintain access, circulated vernacular knowledge, and imagined more nuanced, transparent, and rights-respecting forms of digital governance. Circumvention therefore operated as an everyday form of infrastructural politics, grounded in children's practical engagement with systems that shaped their digital participation. Methodologically, the focus groups made these practices visible by allowing children to explain the social expectations, peer norms, and everyday negotiations surrounding age verification. This supports a child-rights approach in which participation means more than consultation or being heard. It means recognising children as social actors and as genuine experts on their own experiences.

At the same time, sneaking has limits as a metaphor. It captures tactical navigation and informal expertise, but it should not obscure the continuing power asymmetries between children, platforms, and regulators. Platforms still control the infrastructure, even if children find temporary openings within it. Banning children from social media may also restrict access to age-relevant content and reduce opportunities for tailored engagement and parental monitoring, because platforms may design for the assumed absence of child users \autocite{rodriguez2025}. Focus groups also have recognised limitations. Group settings can shape what participants feel comfortable saying, particularly where peer hierarchies, pressure to conform, or sensitive disclosures are involved. Quieter or more marginalised perspectives may therefore be underrepresented. The sample also limits the study's generalisability. Participants were middle-class, urban children, and the small sample size, narrow research timeframe, and qualitative design constrain what can be inferred across socio-economic, cultural, and geographic contexts. The study does not include quantitative measures that might support statistical generalisation. Future research should address these limits by revisiting children's access to restricted social media platforms longitudinally, including parents and developers, and extending participatory co-design work that builds on children's contributions to digital technology policies that directly affect them.

        \section*{Acknowledgements}
We acknowledge the Centre for Artificial Intelligence and Digital Ethics (CAIDE) for its institutional and intellectual support. We also thank the children who participated in this study for their generous contribution of time, experiences, and insights.

\section*{Declarations and Funding Statement}
The authors have no competing interests to declare. This research was supported by the Australian Research Council Future Fellowship Award FT250100270, \emph{Enhancing Ethical Design and Data Use in Child Tracking Apps}, and by Shaanan Cohney's Australian Research Council Discovery Early Career Researcher Award DE260100249, \emph{Safe \& Sound: Privacy-Enhancing Technologies to Protect Young People Online}.

\section*{AI Assistance Statement}
Artificial intelligence tools were used only for editorial assistance after the initial drafting of the manuscript, including cite checking, stylistic suggestions, and typesetting support. The authors reviewed and take full responsibility for all content.

\section*{Ethics}
This research was approved by the University of Melbourne Office of Research Ethics and Integrity. Reference Number: 2026-33398-79843-9.

        \nocite{*}
        \clearpage
        \begingroup
        \small
        \setlength{\bibitemsep}{0.2\baselineskip}
        \printbibliography

@misc{acland2015,
  author = {Acland, C. R. and Dourish, P. and Harris, S. and Holt, J. and Mattern, S. and Miller, T. and colleagues},
  date = {2015},
  title = {Signal Traffic: Critical Studies of Media Infrastructures. University of Illinois Press.}
}

@misc{braunClarke2006,
  author = {Braun, V. and Clarke, V.},
  date = {2006},
  title = {Using thematic analysis in psychology. Qualitative Research in Psychology, 3(2): 77–101.}
}

@misc{bucher2017,
  author = {Bucher, T.},
  date = {2017},
  title = {The algorithmic imaginary: exploring the ordinary affects of Facebook algorithms. Information, Communication \& Society, 20(1): 30-44.}
}

@misc{butler2026,
  author = {Butler, J.},
  date = {2026},
  title = {Two-thirds of under-16s with accounts on Instagram, Snapchat or TikTok kept access despite ban. The Guardian, March 31.}
}

@misc{champion2025,
  author = {Champion, K. E. and Birrell, L. and Smout, S. and Teesson, M. and Slade, T.},
  date = {2025},
  title = {Debate: Social media in children and young people–time for a ban? Beyond the ban–empowering parents and schools to keep adolescents safe on social media. Child and Adolescent Mental Health, 30(4): 411-413.}
}

@misc{chun2011,
  author = {Chun, W. H. K.},
  date = {2011},
  title = {Programmed Visions: Software and Memory. MIT Press.}
}

@misc{coleman2012,
  author = {Coleman, E. G.},
  date = {2012},
  title = {Phreaks, hackers, and trolls. The Social Media Reader, 99-119.}
}

@misc{corbinStrauss1990,
  author = {Corbin, J. and Strauss, A.},
  date = {1990},
  title = {Grounded theory research: Procedures, canons, and evaluative criteria. Qualitative Sociology, 13(1): 3–21.}
}

@misc{deCerteau1984,
  author = {De Certeau, M.},
  date = {1984},
  title = {The Practice of Everyday Life. Trans. Steven Rendall. Berkeley: University of California Press.}
}

@misc{deleuze2017,
  author = {Deleuze, G.},
  date = {2017},
  title = {Postscript on the Societies of Control. In Surveillance, Crime and Social Control. Routledge, 35-39.}
}

@misc{departmentInfrastructure2025a,
  author = {{Department of Infrastructure, Transport, Regional Development, Communications, Sport and the Arts}},
  date = {2025},
  title = {Social Media Minimum Age.}
}

@misc{departmentInfrastructure2025b,
  author = {{Department of Infrastructure, Transport, Regional Development, Communications, Sport and the Arts}},
  date = {2025},
  title = {Age Assurance Technology Trial: Final Report.}
}

@misc{departmentScience2026,
  author = {{Department for Science, Innovation and Technology}},
  date = {2026},
  title = {Growing Up in the Online World: A National Consultation. GOV.UK.}
}

@misc{deuze2011,
  author = {Deuze, M.},
  date = {2011},
  title = {Media Life. Media, Culture \& Society, 33(1): 137-148.}
}

@misc{draperFraser2018,
  author = {Draper, J. T. and Fraser, C. W.},
  date = {2018},
  title = {Beyond The Little Blue Box: The biographical adventures of John T Draper (aka Captain Crunch). Notorious' Phone Phreak', legendary internet pioneer and ardent privacy advocate. Friesen Press.}
}

@misc{esafety2026a,
  author = {{eSafety Commissioner}},
  date = {2026},
  title = {Social Media Age Restrictions.}
}

@misc{esafety2026b,
  author = {{eSafety Commissioner}},
  date = {2026},
  title = {Social Media Minimum Age: March 2026 Compliance Update.}
}

@misc{foucault2018,
  author = {Foucault, M.},
  date = {2018},
  title = {Discipline. In Rethinking the Subject. Routledge, 60-69.}
}

@misc{galloway2004,
  author = {Galloway, A. R.},
  date = {2004},
  title = {Protocol: How Control Exists After Decentralization. MIT Press.}
}

@misc{gallowayThacker2007,
  author = {Galloway, A. R. and Thacker, E.},
  date = {2007},
  title = {The Exploit: A Theory of Networks. University of Minnesota Press.}
}

@misc{gathSwit2024,
  author = {Gath, M. and Swit, C.},
  date = {2024},
  title = {Digital media in early childhood: Risk factors for online harm and psychosocial correlates. Frontiers in Developmental Psychology, 2, Article 1390276.}
}

@misc{gibbs2015,
  author = {Gibbs, M. and Meese, J. and Arnold, M. and Nansen, B. and Carter, M.},
  date = {2015},
  title = {\# Funeral and Instagram: Death, social media, and platform vernacular. Information, Communication \& Society, 18(3): 255-268.}
}

@misc{grother2022,
  author = {Grother, P. and Ngan, M. and Hanaoka, K. and colleagues},
  date = {2022},
  title = {Face Recognition Vendor Test (FRVT): Demographic Summaries. National Institute of Standards and Technology, NIST IR 8429.}
}

@misc{gillespie2018,
  author = {Gillespie, T.},
  date = {2018},
  title = {Custodians of the Internet: Platforms, Content Moderation, and the Hidden Decisions that Shape Social Media. Yale University Press.}
}

@misc{gold2011,
  author = {Gold, S.},
  date = {2011},
  title = {The rebirth of phreaking. Network Security, 2011(6): 15-17.}
}

@misc{hanaoka2024,
  author = {Hanaoka, K. and Ngan, M. and Yang, J. and Quinn G, W. and Hom, A. and Grother, P.},
  date = {2024},
  title = {Face Analysis Technology Evaluation: Age Estimation and Verification. National Institute of Standards and Technology, NIST IR 8525.}
}

@misc{hennessyHeary2005,
  author = {Hennessy, E. and Heary, C.},
  date = {2005},
  title = {Exploring children’s views through focus groups. Researching Children’s Experience: Approaches and Methods, 236-252.}
}

@misc{hfAn2025,
  author = {HF, N. H. and An Nur, F.},
  date = {2025},
  title = {Constructing childhood realities: A communication science analysis of TikTok’s age verification failure and psychosocial impact. Legal Brief, 14(3).}
}

@misc{hourcade2023,
  author = {Hourcade, J. P. and Alper, M. and Antle, A. and colleagues},
  date = {2023},
  title = {Developing participatory methods to consider the ethics of emerging technologies for children. Proceedings of CHI, 1-3.}
}

@misc{howard2025,
  author = {Howard, F. and Hanckel, B. and Moore, K. and Atherton, S. and Suppers, J.},
  date = {2025},
  title = {Researching with Young People: An Introduction to Youth-centred Research Methods. Bloomsbury Publishing.}
}

@misc{onlineSafety2024,
  author = {{Online Safety Amendment (Social Media Minimum Age) Act}},
  date = {2024},
  title = {(Cth).}
}

@misc{kittler1999,
  author = {Kittler, F. A.},
  date = {1999},
  title = {Gramophone, Film, Typewriter (G Winthrop‑Young M Wutz, Trans). Stanford University Press.}
}

@misc{kitzinger1994,
  author = {Kitzinger, J.},
  date = {1994},
  title = {The methodology of focus groups: the importance of interaction between research participants. Sociology of Health \& Illness, 16(1): 103-121.}
}

@misc{kohler2025,
  author = {Köhler-Dauner, F. and Peter, L. and Sitarski, E. and Chauviré-Geib, K. and Haag, A. -. C. and Fegert, J. M.},
  date = {2025},
  title = {Digital child protection in social networks: Age verification and age-tiered regulation in Europe Child and Adolescent Psychiatry and Mental Health, 19, Article 143.}
}

@misc{lakoffJohnson1980,
  author = {Lakoff, G. and Johnson, M.},
  date = {1980},
  title = {Metaphors We Live By University of Chicago Press.}
}

@misc{lavoipierre2025,
  author = {Lavoipierre, A.},
  date = {2025},
  title = {VPNs, ``old man'' masks, and AI: The holes in the social media ban and their fixes. ABC News, October 1.}
}

@misc{levy1984,
  author = {Levy, S.},
  date = {1984},
  title = {Hackers: Heroes of the Computer Revolution. Anchor Press/Doubleday.}
}

@misc{livingstoneOneill2014,
  author = {Livingstone, S. and O'Neill, B.},
  date = {2014},
  title = {Children’s rights online: Challenges, dilemmas and emerging directions. In S van der Hof, B van den Berg, B Schermer (Eds), Minding Minors Wandering the Web: Regulating Online Child Safety. Springer, 19-38.}
}

@misc{livingstoneThird2017,
  author = {Livingstone, S. and Third, A.},
  date = {2017},
  title = {Children and young people’s rights in the digital age: An emerging agenda. New Media \& Society, 19(5): 657-670.}
}

@misc{ludlow2022,
  author = {Ludlow, P.},
  date = {2022},
  title = {Science, Metaphors, and Memes. In: Wuppuluri, S, Grayling, AC (eds) Metaphors and Analogies in Sciences and Humanities. Synthese Library, vol 453. Springer, Cham.}
}

@misc{lueks2026,
  author = {Lueks, W. and Gurses, S. and Diaz, C.},
  date = {2026},
  title = {Assessing age assurance technologies: Effectiveness, side-effects, and acceptance. arXiv.}
}

@misc{morgan1997,
  author = {Morgan, D. L.},
  date = {1997},
  title = {Focus Groups as Qualitative Research (Vol. 16). Sage.}
}

@misc{pasquale2015,
  author = {Pasquale, F.},
  date = {2015},
  title = {The Black Box Society: The Secret Algorithms that Control Money and Information Harvard University Press.}
}

@misc{plantinPunathambekar2019,
  author = {Plantin, J. C. and Punathambekar, A.},
  date = {2019},
  title = {Digital media infrastructures: Pipes, platforms, and politics. Media, Culture \& Society, 41(2): 163-174.}
}

@misc{reuters2026a,
  author = {{Reuters}},
  date = {2026},
  title = {French Senate debates social media ban for children under 15. Reuters.}
}

@misc{reuters2026b,
  author = {{Reuters}},
  date = {2026},
  title = {Germany’s ruling party backs social media curbs for children. Reuters.}
}

@misc{rodriguez2025,
  author = {Rodriguez, A. and Dezuanni, M. and Heck, E.},
  date = {2025},
  title = {Demystifying Australia’s Teen Social Media Ban. DMRC.}
}

@misc{rubinstein2024smma,
  author = {Rubinstein, B. I. P. and Ohrimenko, O. and Cullen, A. C. and Cohney, S. and Culnane, C. and Murray, T. and Cheong, M. and Pham, T. and Yuan, X.},
  date = {2024},
  title = {Submission 272 to the Senate Environment and Communications Legislation Committee: Online Safety Amendment (Social Media Minimum Age) Bill 2024.}
}

@misc{sandbergGille2021,
  author = {Sandberg, H. and Gille, J.},
  date = {2021},
  title = {Investigating the digital media engagements of very young children at home: Reflecting on methodology and ethics. Communications, 46(3): 332-351.}
}

@misc{smahel2020,
  author = {Smahel, D. and Machackova, H. and Mascheroni, G. and Dedkova, L. and Staksrud, E. and Ólafsson, K. and Livingstone, S. and Hasebrink, U.},
  date = {2020},
  title = {EU Kids Online 2020: Survey Results From 19 Countries. EU Kids Online.}
}

@misc{srnicek2017,
  author = {Srnicek, N.},
  date = {2017},
  title = {Platform Capitalism. John Wiley \& Sons.}
}

@misc{stalfordLundy2025,
  author = {Stalford, H. and Lundy, L.},
  date = {2025},
  title = {Whose business? Protecting children’s rights in the online environment. The International Journal of Children's Rights, 33(1): 1-4.}
}

@misc{stoilova2021,
  author = {Stoilova, M. and Nandagiri, R. and Livingstone, S.},
  date = {2021},
  title = {Children’s understanding of personal data and privacy online: A systematic evidence mapping. Information, Communication \& Society, 24(4): 557-575.}
}

@misc{taylor2026,
  author = {Taylor, J.},
  date = {2026},
  title = {Australia wants to sell its social media ban to the world – but are the measures even working? The Guardian, April 1.}
}

@misc{unitedNations1989,
  author = {{United Nations}},
  date = {1989},
  title = {Convention on the Rights of the Child.}
}

@misc{unCommittee2021,
  author = {{UN Committee on the Rights of the Child}},
  date = {2021},
  title = {General comment No. 25 (2021) on children’s rights in relation to the digital environment.}
}

@misc{valguarneraLandoni2023,
  author = {Valguarnera, S. and Landoni, M.},
  date = {2023},
  title = {Design with and for children: The challenge of inclusivity. In Proceedings of CHI, 171-184.}
}

@misc{warnick2004,
  author = {Warnick, B. R.},
  date = {2004},
  title = {Technological metaphors and moral education: The hacker ethic and the computational experience. Studies in Philosophy and Education, 23(4): 265-281.}
}

@misc{ytreMoe2021,
  author = {Ytre-Arne, B. and Moe, H.},
  date = {2021},
  title = {Folk theories of algorithms: Understanding digital irritation. Media, Culture \& Society, 43(5): 807-824.}
}
        \endgroup
        \end{document}